# Oscillating Magnetic Effect in BiFeO$_3$

Thiago Ferro[1,2], Adrielson Dias[2,3], Maria Clara[2,3], Luana Hildever[1,2], and José Holanda[1,2,3*]

[1]Programa de Pós-Graduação em Engenharia Física, Universidade Federal Rural de Pernambuco, 54518-430, Cabo de Santo Agostinho, Pernambuco, Brazil

[2]Group of Optoelectronics and Spintronics, Universidade Federal Rural de Pernambuco, 54518-430, Cabo de Santo Agostinho, Pernambuco, Brazil

[3]Unidade Acadêmica do Cabo de Santo Agostinho, Universidade Federal Rural de Pernambuco, 54518-430, Cabo de Santo Agostinho, Pernambuco, Brazil

**Abstract**

The development of electric vehicles has led to a growing need for more efficient and environmentally friendly batteries. As a result, there is significant interest in researching new materials and techniques to enhance battery efficiency. One such material being explored is bismuth ferrite (BiFeO$_3$ or BFO), a perovskite with versatile properties. Researchers are particularly intrigued by the potential to control its antiferromagnetic magnetization using magnetic or electric fields. Here, a comprehensive analysis of BFO was conducted, with a focus on its behavior when subjected to oscillating magnetic fields. The research revealed that BFO is sensitive to the frequency and shape of these magnetic fields, leading to the discovery of a new effect related to the transmission of electromagnetic signals on its surface. This effect resulted in a significant increase in the power of the electromagnetic signal, representing a major technological breakthrough. According to the findings, this gain in power has not been observed in any system of this kind before. The study also demonstrated that BFO has the ability to detect magnetic fields through electrical output signals and vice versa, which is crucial for assessing the state and efficiency of batteries, thus contributing to significant advancements in energy storage technology.

Corresponding author: * joseholanda.silvajunior@ufrpe.br

# 1. Introduction

The increasing demand for electrical energy in technologies such as portable electronic devices, renewable energy storage systems, and electric vehicles has led to a need for more efficient and safe batteries. Batteries are crucial for modern life, enabling the storage, consumption, and transportation of energy. It is projected that the battery market will grow by $330 billion by 2032, compared to 2022, and the installed costs of battery storage systems are expected to decrease by 50% to 66% by 2030 due to advancements in performance and the use of new materials. Multiferroic materials, which possess multiple ferroic orders that can be controlled separately or simultaneously, are being considered for energy storage applications. However, only a limited number of these materials exhibit magnetoelectric coupling at room temperature. Bismuth ferrite ($BiFeO_3$ or BFO) is one such material and is considered a promising candidate for technological applications, particularly in the energy sector.

BFO has a remanent polarization of approximately 100μC/cm$^2$, making it the perovskite with the highest switchable polarization among perovskite ferroelectrics. It is also lead-free and has the potential to be used in ferroelectric devices and memories. Additionally, BFO's magnetoelectric coupling allows for the detection of magnetic fields through electrical output signals and vice versa, which is crucial for assessing the state and efficiency of batteries. BFO is an antiferromagnetic perovskite with a rhombohedral crystal structure and exhibits a spontaneous antiferromagnetic magnetization due to G-type antiferromagnetic coupling between the $Fe^{3+}$ sites. The ferroelectric phase corresponds to the symmetry of the R3c space group with cations displaced in the $[111]_C$ direction with tilted oxygen octahedra. BFO has an antiferromagnetic transition temperature ($T_N$) of 643 K and a ferroelectric transition temperature ($T_C$) of 1098 K. In this work, BFO was analyzed using oscillating magnetic fields, revealing its sensitivity to the frequency and shape of these fields. This led to the discovery of a new effect during the transmission of electromagnetic signals on its surface. This discovery has the potential for application in energy storage technology and opens new possibilities for the development of oscillating magnetic field sensors to enhance battery efficiency and safety.

## 2. Experimental procedures

### (a) Characterization

Our samples underwent X-ray diffraction (XRD) analysis using Cu-Kα radiation with a wavelength of 0.154 nm, and in the Bragg angle range of 2θ from 20° to 80°. In **Figure 1(a)**, a typical X-ray diffraction measurement of our samples is shown. The XRD peaks indicate a single-phase rhombohedral perovskite with the space group R3c (ICSD 86-1518). The sharp XRD peaks demonstrate the crystallinity of our samples. Comparison with ICSD confirmed small traces of the ferrimagnetic phase $Fe_3O_4$, as observed in ceramics with compositions similar to the one we studied here [23–25]. Analysis of the average crystallite size using the Debye–Scherrer formula yielded a value of 31.9 nm. The XRD patterns were refined using the Rietweld refinement method with PowderCell to obtain lattice parameters, rhombohedral unit cell volume, Fe–O–Fe bond angle, Fe–O bond length, and Bi–O bond length, which are essential for understanding the properties exhibited in the R3c structure [26–28]. The refined results are provided in **Table 1**.

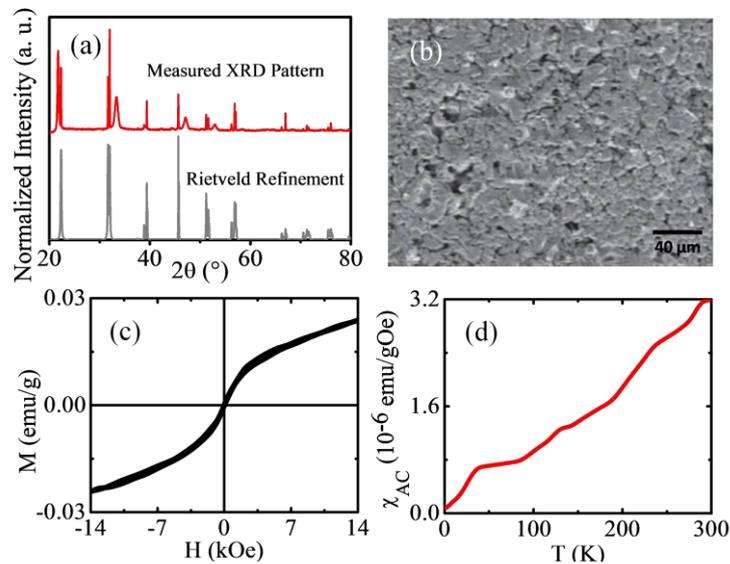

**Figure 1:** **(a)** X-ray diffraction (XRD) patterns of bismuth ferrite ($BiFeO_3$ or BFO), **(b)** typical micrograph of the topology of our samples obtained using a 200 kV FEG Quanta scanning electron microscope, **(c)** Magnetic hysteresis loops typical of the samples, and **(d)** T-dependence for $\chi_{AC}$ measured by using an AC magnetic field with amplitude of 15 Oe and frequency of 1 kHz.

**Table 1:** The refined lattice parameters, structure fitting factors, Fe–O–Fe bond angle, Fe–O bond length and Bi–O bond length of bismuth ferrite (BiFeO$_3$ ou BFO).

| Sample | BFO |
|---|---|
| Lattice parameters (nm) | $a = 0.5579$, $c = 1.3869$ |
| V (nm$^3$) | 0.3738 |
| R-factors (%) | $R_p = 8.48$, $R_{exp} = 4.25$ |
| $\chi^2$ | 4.48 |
| Fe-O-Fe bond angle (°) | 153.3717 |
| Fe-O bond length (nm) | 1.9787 |
| Bi-O bond length (nm) | 2.4726 |

We conducted a topological analysis of our samples using a 200 kV FEG Quanta scanning electron microscope. We observed that the topology is of good quality as the samples are ceramic. In **Figure 1(b)**, we display a typical micrograph of the samples' topology. In **Figure 1(c)**, we present a typical hysteresis loop of the samples obtained in a magnetic field range of ±15 kOe. Analyzing the hysteresis loop, we obtained a saturation magnetization of 0.00244 emu/g, a remanent magnetization of 0.0241 emu/g, and a coercivity of 530 Oe. In **Figure 1(d)**, we show the AC magnetic susceptibility, $\chi_{AC} = \sqrt{\chi_R^2 + \chi_I^2}$, where $\chi_R$ and $\chi_I$ are the in-phase and out-of-phase components of $\chi_{AC}$, respectively, for a frequency of 1 kHz, AC magnetic field magnitude of 15 Oe, and temperatures between 5 and 300 K. These results are in good agreement with references [26] to [31] [26-31].

## (b) Experimental Setup

*Experimental*

We conducted experiments using pulsed electromagnetic waves in the radio-frequency range under oscillating magnetic fields in three different configurations, as depicted in **Figure 2**. The pulses travel along the sample surface through an emitting antenna positioned at one end of the sample, and are received by another antenna located at the opposite end. The direction of pulse propagation is always aligned with the long dimension of the sample. We utilized pulse widths of 5, 10, and 15 μs with an input power of 10 mW. In **Figure 2(a)**, the pulsed wave propagation occurs perpendicular to the oscillating and static magnetic fields. In **Figure 2(b)**, the pulsed wave propagation is parallel to the oscillating magnetic fields and perpendicular to the static magnetic fields. Lastly, in **Figure 2(c)**, the pulsed wave propagation is in the direction of the static magnetic fields and perpendicular to the oscillating magnetic fields.

*Simulated*

We conducted Python simulations of pulsed electromagnetic waves using the Heaviside function to improve the accuracy of the pulse description. Additionally, we utilized the Damon-Eshbach equation [32] to simulate the magnetostatic modes permitted in our samples, using the parameters obtained during the characterization process. The goal of these analyses was to determine the dispersion curves for the two types of modes (surface and bulk) present in our samples.

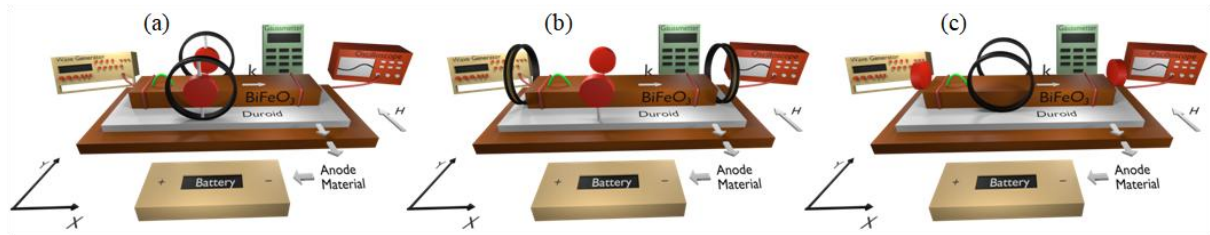

**Figure 2**: Pulses are transmitted across the sample surface using an emitting antenna positioned at one end of the sample and received by another antenna located at the opposite end. The pulses can propagate in three different ways: **(a)** in a direction perpendicular to the oscillating and static magnetic fields, **(b)** in a direction parallel to the oscillating magnetic fields and perpendicular to the static magnetic fields, and **(c)** in a direction parallel to the static magnetic fields and perpendicular to the oscillating magnetic fields.

## 3. Results and discussions

In **Figure 3**, we provide a description of the pulses transmitted on the material surface via the antenna, as well as the pulses simulated using the Heaviside function. The striking similarity between the experimental and simulated pulses indicates that it will be possible to compare the experimental and simulated results for the power as a function of the oscillating field applied to the material. **Figures 3 (a)**, **(b)**, and **(c)** show the experimental pulses with an output power of 10 mW and pulse widths of 5, 10, and 15 µs with a period of 2 ms. Similarly, **Figures 3 (d)**, **(e)**, and **(f)** display the simulated pulses using the Heaviside function with an output amplitude of 10 mW and pulse widths of 5, 10, and 15 µs with a period also of 2 ms. Additionally, in **Figures 3 (g)**, **(h)**, and **(i)**, a pulse with an output power of 10 mW, width of 10 µs, period of 2 ms is shown without (red signals) and with (green signals) oscillating magnetic fields of amplitude of 100 Oe and static magnetic fields of 10 kOe. These figures illustrate the pulsed wave propagation in different directions relative to the oscillating and static magnetic fields.

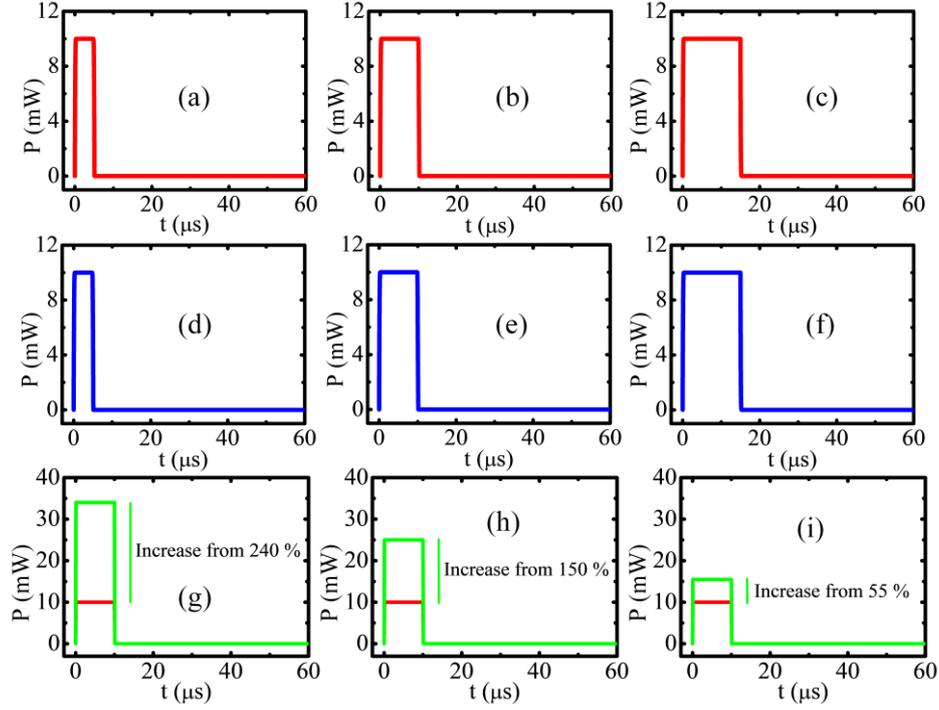

**Figure 3:** Experimental pulses were transmitted through an antenna inserted on the surface and end of the sample. Each pulse had an output power of 10 mW, a period of 2 ms, and no applied magnetic fields. The pulse widths were 5 μs **(a)**, 10 μs **(b)**, and 15 μs **(c)**. Simulated pulses were created using the Heaviside function with an output amplitude of 10 mW and pulse widths of 5 μs **(d)**, 10 μs **(e)**, and 15 μs **(f)** with a period of 2 ms. There was also a pulse with an output power of 10 mW, a width of 10 μs, and a period of 2 ms, without (red signals) and with (green signals) oscillating magnetic fields of amplitude 100 Oe and static magnetic fields of 10 kOe. The pulses were propagated in different directions in relation to the magnetic fields: **(g)** pulsed wave propagated perpendicular to the oscillating and static magnetic fields, **(h)** pulsed wave propagated parallel to the oscillating magnetic fields and perpendicular to the static magnetic fields, and **(i)** pulsed wave propagated in the direction of the static magnetic fields and perpendicular to the oscillating magnetic fields.

In **Figures 4 (a)**, **(b)**, and **(c)**, we display the results obtained using the Damon-Eshbach equation [32] for the dispersion relation. These results correspond to the three configurations shown in **Figure 2**. The first configuration involves the pulsed wave propagating perpendicular to the oscillating and static magnetic fields. In the second configuration, the pulsed wave propagates parallel to the oscillating magnetic fields and perpendicular to the static magnetic fields. Finally, the third configuration entails the pulsed wave propagating in the direction of the static magnetic fields and perpendicular to the oscillating magnetic fields. During these simulations, the oscillating magnetic field had an amplitude of 100 Oe, while the static magnetic field was 10 kOe. In **Figures 4 (d)**, **(e)**, and **(f)**, we present the results of the experimental measurements obtained for each configuration in **Figure 2**. These experimental measurements were conducted with an oscillating magnetic field amplitude of 100 Oe and a static magnetic field of 10 kOe. The solid lines in **Figures 4 (d)**, **(e)** and **(f)** represent the power generated by the oscillating fields for the three configurations presented in **Figure 2**, which were obtained through the equation $P(f) = 2Zd^2(f/\gamma)^2$ [32, 33], where $Z = 50\ \Omega$ is the system impedance, $d = 5$ mm is the sample width, and $f$ is the frequency with $\gamma = 2.8$ MHz/Oe being the gyromagnetic factor. In addition, we calculated the magnetic field $h_y$ due to the output electromagnetic current $i_{ele} = 20$ mA, for each wave vector $k_y$, i.e., $h_y = \pi(i_{ele})k_y$ [26-33].

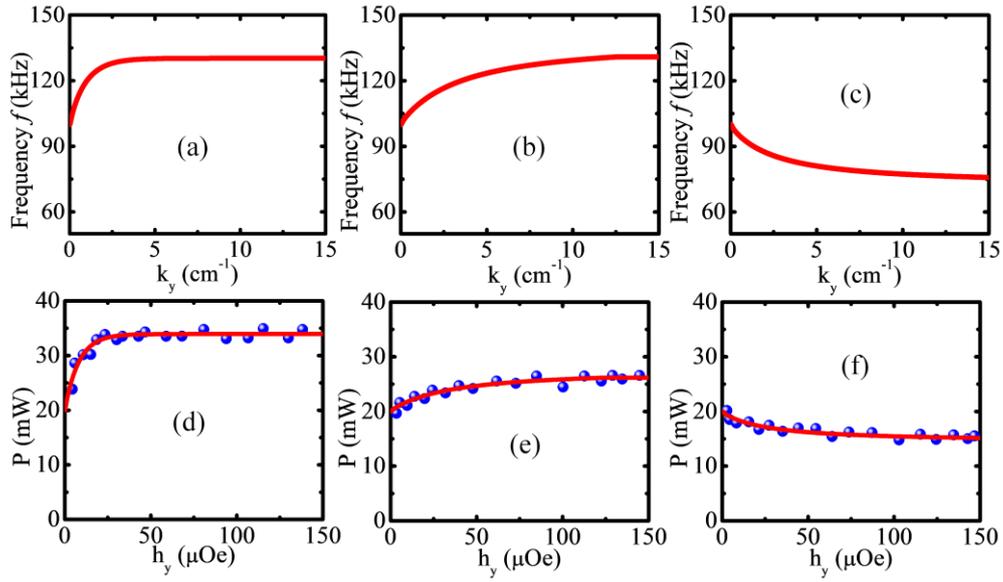

**Figure 4:** Simulations were conducted using the Damon-Eshbach equation [32] to determine the dispersion relation for the propagation of the pulsed wave in the following scenarios: **(a)** perpendicular to the oscillating and static magnetic fields, **(b)** parallel to the oscillating magnetic fields and perpendicular to the static magnetic fields, and **(c)** in the direction of the static magnetic fields and perpendicular to the oscillating magnetic fields. Experiments were carried out to measure the powers as a function of the oscillating magnetic fields in the following scenarios: **(d)** perpendicular to the oscillating and static magnetic fields, **(e)** parallel to the oscillating magnetic fields and perpendicular to the static magnetic fields, and **(f)** in the direction of the static magnetic fields and perpendicular to the oscillating magnetic fields. The solid lines in items **(d)**, **(e)** and **(f)** represent the respective powers obtained through the equation $P(f) = 2Zd^2(f/\gamma)^2$ [32, 33], where $Z = 50\ \Omega$ is the system impedance, $d = 5$ mm is the sample width, and $f$ is the frequency with $\gamma = 2.8$ MHz/Oe being the gyromagnetic factor. In addition, we calculate the magnetic field $h_y$ due to the output electromagnetic current $i_{ele} = 20$ mA, for each wave vector $k_y$, that is, $h_y = \pi(i_{ele})k_y$ [26-33].

## Conclusion

In our observations, as depicted in **Figures 4 (d)**, **(e)**, and **(f)**, we found that the most effective configuration for enhancing power through the oscillating magnetic effect is that of **Figure 2 (a)**. **Figure 3** shows that for a field of 100 μOe, the gain can reach up to 240% for the configuration of **Figure 2 (a)**, 150% for **Figure 2 (b)**, and 55% for **Figure 2 (c)**. We also noted that the oscillating magnetic effect is sensitive to frequency and can be explained by the dispersion relations identified using the Damon-Eshbach approach [32]. This effect, in magnetic terms, originates from the uncompensated and random spins on the surface of our samples, indicating that it is a local effect that, when appropriately excited, becomes a global effect throughout the sample, contributing to the overall increase in power transmitted through the material. We believe that this observation holds significant importance for battery applications, as it provides a means to amplify the power of an electromagnetic signal and, consequently, the energy transmitted by the system.

## Acknowledgements

This research was supported by Conselho Nacional de Desenvolvimento Científico e Tecnológico (CNPq) with Grant Number: 309982/2021-9, Coordenação de Aperfeiçoamento de Pessoal de Nível Superior (CAPES) with Grant Number: PROAP2023UFRPE, and Fundação de Amparo à Ciência e Tecnologia do Estado de Pernambuco (FACEPE) with Grant Number: IBPG-1292-3.03/22. The authors are grateful to professor Francisco Estrada of Universidad Michoacana de San Nicolas de Hidalgo and professor Changjiang Liu of the Department of Physics of The State University of New York at Buffalo (University at Buffalo) for the valuable discussions on this research, and to the Centro Multiusuário de Pesquisa e Caracterização de Materiais da Universidade Federal Rural de Pernambuco (CEMUPEC-UFRPE). Adrielson de Araújo Dias (Adrielson Dias) and Maria Clara Gonçalves Santos (Maria Clara) acknowledge the Program for Scientific Initiation Scholarships (PIBIC/PIC) of the CNPq/UFRPE.

## Contributions

T. F., A. D., M. C., and L. H. analyzed all the experimental measures and J. H. discussed, wrote and supervised the work.

## Conflict of interest

The authors declare that they have no conflict of interest.

**Data availability statement**

The data generated and/or analysed during the current study are not publicly available for legal/ethical reasons but are available from the corresponding author on reasonable request.